\documentclass[12pt,a4paper,twoside]{article}        
\usepackage[]{babel}                  
\usepackage[cp1251]{inputenc}                        
\usepackage{mmgart}                                  
\begin{document}

\setcounter{page}{13}                                
\thispagestyle{empty}                                
\begin{heading}                                      
{Volume\;5,\, N{o}\;2,\, p.\,1 -- 11\, (2017)}      
{Special issue}                                      
\end{heading}                                        

\begin{Title}
Faddeev calculations for light $\Xi$-hypernuclei 
\end{Title}

\begin{center}
\Author{a}{Igor Filikhin},
\Author{}{Vladimir M. Suslov}
\and
\Author{}{Branislav Vlahovic}
\end{center}


\begin{flushleft}
\Address{}{Department of Mathematics and Physics, North Carolina Central University, 1801 Fayetteville Street, Durham, NC 27707, USA
}

\Email{$^a$\,ifilikhin@nccu.edu}
\end{flushleft}

\Headers{I. Filikhin et al}{
 Light $\Xi$ hypernuclei 
}

\begin{flushleft}                                 
\small\it Received 19 April 2017, 
\ Published 3 May 2017.        
\end{flushleft}                                   

\Thanks{
This work is supported by the NSF (HRD-1345219)  and NASA (NNX09AV07A)
}

\Thanks{\mbox{}\\
\copyright\,The author(s) 2017. \ Published by Tver State University, Tver, Russia}
\renewcommand{\thefootnote}{\arabic{footnote}}
\setcounter{footnote}{0}

\Abstract{
The  hypernuclear systems $NN\Xi $ and $\Xi\Xi N$ are considered as an analogue of $nnp$ ($^3$H) nuclear system (with the notation as $AAB$ system). We use the recently proposed modification for the $s$-wave Malfliet-Tjon potential. The modification simulates 
 the Extended-Soft-Core model (ESC08c) for baryon-baryon interactions. The  $\Xi N$ spin/isospin triplet $(S, I)=(1, 1)$ potential generates  a bound state  with  the energy $B_2(AB)$=1.56~MeV.  
Three-body binding energy $B_3$ for the states with maximal total isospin  is calculated employing the configuration-space Faddeev equations. 
Comparison with the results obtained  within the integral representation for the equations is presented.
The different types of the relation between $B_2$ and $B_3(V_{AA}=0)$ are discussed.
 }
\\
\noindent
\Keywords{Few-body systems, Hypernuclei, Nuclear forces, Faddeev equations}\\
\PACS{21.45.-v, 21.80.+a, 21.30.Fe, 11.80.Jy}


\newpage                               
\renewcommand{\baselinestretch}{1.1}   

\section{Introduction}
The first $\Xi$-hypernuclear bound state has been reported in  Ref. \cite{WLRL}. The lifetime of a $\Xi$-hypernucleus is long
enough to enable the hypernuclear state to be
well defined.  According the  current experimental
data, the $\Xi$-nucleus interactions are attractive \cite{N}.
In particular, the hyprnucleus  $^{12}_\Xi$Be can be interpreted by assuming
a nucleus Wood-Saxon potential with a strength parameter of about -14~MeV \cite{Kha}.
Another  hypernucleus $^{15}_\Xi$C is considered to be
the cluster system $^{14}$N(ground state) + $\Xi$, where $\Xi$ can be in $s$ or $p$-wave state \cite{SHSSM}.

The stable states in the systems $\Xi NN$ and $\Xi\Xi N$  were recently predicted in Refs. \cite{GV15,GV16,GVV16} based on the recent update of  the Extended-Soft-Core (ESC08c) model  
\cite{NRY15, RS16, RNY} for baryon-baryon interactions.  This model has predicted the $\Xi N$  bound spin/isospin triplet $(S, I)=(1, 1)$ state with three-body energy $B_3$ to be equal to 1.56~MeV. This bound state of
proton and $\Xi^0$ or neutron and $\Xi^-$ has  maximal isospin of the $\Xi N$ pair.
For the three-body systems when all  pairs $NN$, $\Xi N$, and $\Xi\Xi$ are in triplet isospin states,
the strong decay
$N\Xi \to \Lambda\Lambda$ is forbidden. Such three-body systems can 
 be stable under the strong interaction.
 The first calculations \cite{GV16,GVV16} based on the assumption yield the existence of bound states for 
the  $\Xi NN$ and $\Xi\Xi N$ systems.

 In the presented work, we use the differential  Faddeev equations to  mathematically formulate the 
 bound state problems for the $\Xi NN$ and $\Xi\Xi N$ systems.
 The alternative treatment is presented in Refs. \cite{GV16,GVV16} where  the integral Faddeev equations were applied.
  Our calculations for the systems are generally in the agreement with the results \cite{GV16,GVV16}. However, we found that a small correction for the results  is needed. 
We present our results along with the  correction \cite{G16} of the results published in Refs. \cite{GV16,GVV16}.
Additionally, the binding energy for the spin, isospin ($0$, $1$) bound state for the $\Xi \Xi \alpha$ system is calculated.   This state was not considered in Refs. \cite{GV16,GVV16}. 

  The models for $\Xi NN$ and $\Xi\Xi N$ ($\Xi \Xi \alpha$) are restricted by the $s$-wave approach.
 The coupling to higher-mass channels, $\Sigma\Lambda$
and $\Sigma\Sigma$, does not taken into account assuming that their contributions  have the second order of smallness
to the binding energy of three-body system. The calculations do not also take into account the Coulomb force.

\section{Formalism}
\subsection{Faddeev equations for AAB system}
The differential Faddeev equations  \cite{MF85} can be reduced to a simpler form for the case of two identical particles (like an $AAB$ system). In this case the total wave function of the system is decomposed into the sum of the Faddeev components $U$ and $W$ corresponding to the $(AA)B$ and $(AB)B$ types of rearrangements: $\Psi =U+W\pm PW$, where $P$ is the permutation operator for two identical particles. In the latter expression the sign ''$+$'' corresponds to two identical bosons, while the sign ''$- $'' corresponds to two identical fermions, respectively. The
set of the Faddeev equations is written as following: 
\begin{equation}
\begin{array}{l}
{(H_{0}+V_{AA}-E)U=-V_{AA}(W\pm PW)}, \\ 
{(H_{0}+V_{AB}-E)W=-V_{AB}(U \pm PW)}.
\end{array}
\label{GrindEQ__1_}
\end{equation}%
Here, $H_{0}$ is the operator of kinetic energy of the Hamiltonian taken for corresponding Jacobi coordinates. The functions $V_{AA}$ and $V_{AB}$ describe the pair interactions between the particles.
The model space is restricted to the states with the total angular momentum $L=0$, the momentum of pair $l=0$, and momentum  $\lambda=0$ of the third particle  respectively to the center of mass of the pair. 

\subsection{$s$-wave approach}
The description of the above mentioned $AAB$ systems is distinguished by the masses of particles and the type of ${AA}$ and ${AB}$ interactions. We use  $s$-wave $V_{AA}$ and $V_{AB}$ potentials, which are spin-isospin dependent. This requires to write Eq. (\ref{GrindEQ__1_}) with  the corresponding spin-isospin configurations. 

The separation of spin-isospin variables leads to the Faddeev equations for the considering systems in the following form: 
\begin{equation} 
\label{eq:1} 
\begin{array}{l} 
   (H_0+V_{AA}-E){\cal U}
  =-V_{AA}D(1+p){\cal W} \;\; , \\
   (H_0+V_{AB}-E){\cal W}=-V_{AB} 
  (D^T{\cal U}+Gp{\cal W}) \;\; , 
\end{array} 
\end{equation} 
where matrices $D$ and $G$ are defined by the nuclear system under consideration, ${\cal W}$ is a column matrix with the singlet and triplet parts of the $W$ component of the wave function of a nuclear system, and the 
exchange operator $p$ acts on the coordinates of identical particles. 

For the $^3$H nucleus, considered as $pnn$ system in the state $(S,I)$=$(1/2, -)$, 
we applied the isospin-less approach proposed in Ref. \cite{FSV16}.
The inputs into Eq. (\ref{eq:1}) are the following: the spin singlet $nn$ 
potential $V_{AA}=v^s_{nn}$ and $V_{AB}=diag\{v^s_{np},v^t_{np}\}$ that
is a diagonal $2\times2$ matrix with the spin singlet $v^s_{np}$ and spin triplet $v^t_{np}$ 
 $np$ potentials, respectively, 
and 
\begin{equation}
\label{w1}
D=(-\frac12,\frac{\sqrt3}2),   \quad
G=\left( \begin{array}{rr}
-\frac12& -\frac{\sqrt3}2 \\
-\frac{\sqrt3}2& \frac12 \\
\end{array} \right),  \quad
{\cal W}=\left( \begin{array}{rr}
{\cal W}^{s} \\ 
{\cal W}^{t} \\ 
\end{array} \right), \quad {\cal U}={\cal U}^{s},
\end{equation}
where ${\cal W}^{s}$ and ${\cal W}^{t}$ are the spin singlet and spin triplet parts of the ${\cal W}$ component.
Within the isospin formalism when the protons and neutron are identical particles, instead of Eq. (\ref{eq:1}), which is a set of three equations, one has the set of two equations for the state $(S,I)$=$(1/2, 1/2)$ of  the three nucleon  system $NNN$:
\begin{equation} 
\label{eq:1s} 
\begin{array}{l} 
    (H_0+V_{NN}-E){\Phi}=-V_{NN} {B}(p^++p^-){\Phi} \;\; , 
\end{array} 
\end{equation} 
where 
$$ 
\label{eq:matrix1} 
V_{NN}=diag\{v^s_{NN},v^t_{NN}\}, \qquad
{ B}=\left( \begin{array}{rr}
 \frac14& -\frac34 \\
-\frac34&  \frac14 \\
\end{array} \right), \qquad 
{\Phi}=\left( \begin{array}{ll}
{\Phi}^{s} \\ 
{\Phi}^{t} \\ 
\end{array} \right)
$$
and $p^{\pm}$ are the operators of cyclical permutations  for coordinates of the particles. 

\subsection{Spin-isospin configurations} 

\begin{figure}[!ht]
\centering
\includegraphics[width=8pc]{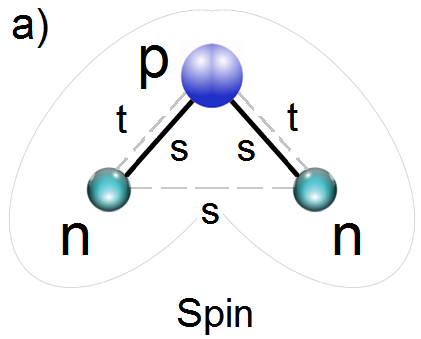}
\includegraphics[width=16pc]{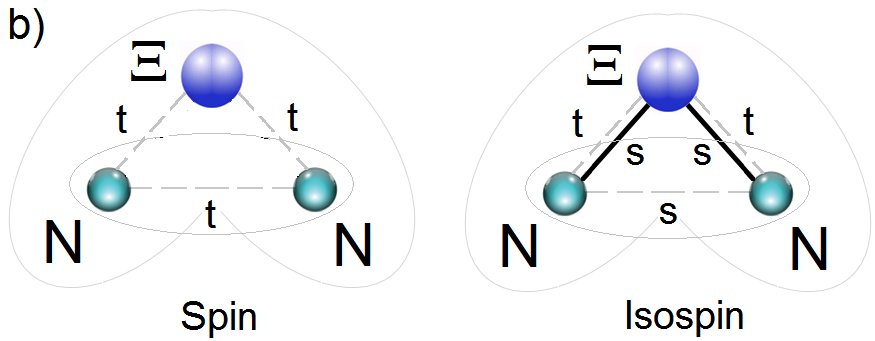}
\includegraphics[width=16pc]{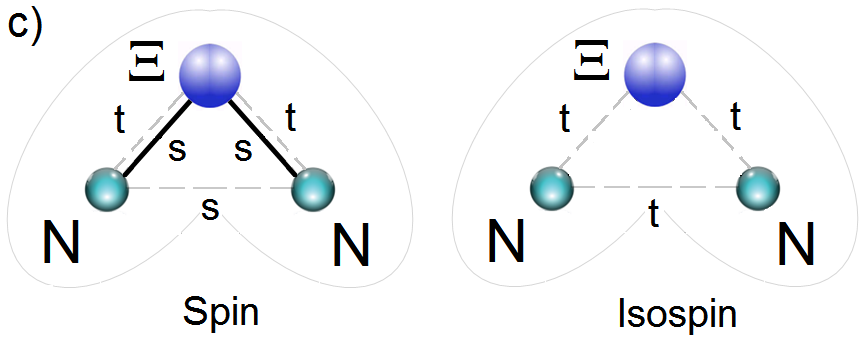}
\includegraphics[width=16pc]{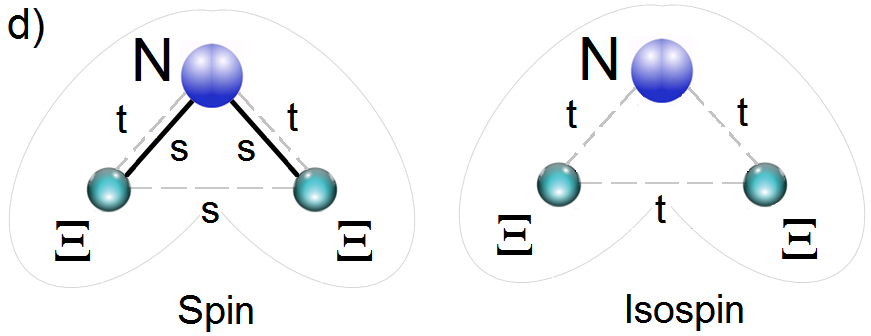}
\includegraphics[width=8pc]{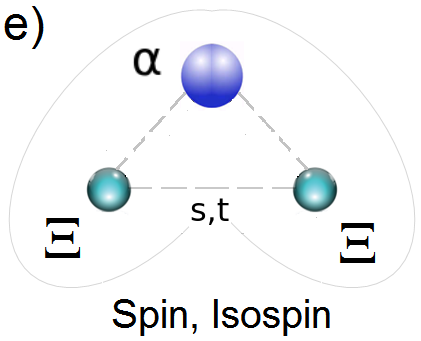}
\caption{\label{fig3} \small The spin-isospin configurations in $AAB$  systems: a) $nnp$, $(S,I)$= $(1/2,-)$,    b) $NN{\Xi}$,   $(S,I)$=$(3/2,3/2)$, c) $NN{\Xi}$,   $(S,I)$=$(1/2,3/2)$, d) ${\Xi}{\Xi}N$,  $(S,I)$=$(1/2,3/2)$, e)  ${\Xi}{\Xi}\alpha$,  $(S,I)$=$(0,1)$. 
The pair  potentials have spin or isospin singlet and triplet components (noted as $s$ and $t$).  The  two-body bound  states are noted by ovals.
}
\end{figure}

In Eq. (\ref{GrindEQ__1_}), the Faddeev  component $U$ (and $W$) of the total wave-function  is  expressed in terms of spin and isospin:
$$
U={\cal U} \chi_{spin}\eta_{isospin}.
$$
The graphical representation of the spin-isospin configurations in the ${\Xi}{\Xi}N$ and  $NN{\Xi}$ systems is given in Fig. \ref{fig3}. 
Here, we have taken into account that the spin (isospin) basis of the spin (isospin) $3/2$ state for three-body $AAB$ system
is formed by a single basis element. Thus, the Faddeev equations for each system considered have the form   (\ref{eq:1})-(\ref{w1}).
The equation for the state ${\Xi}{\Xi}\alpha$ $(S,I)$=$(0,1)$ has a "scalar form" instead the form  (\ref{eq:1})-(\ref{w1}):
\begin{equation} 
\label{eq:1av} 
\begin{array}{l} 
   (H_0+V_{AA}-E){\cal U}
  =-V_{AA}(1+p){\cal W} , \\
   (H_0+V_{AB}-E){\cal W}=-V_{AB}   ({\cal U}+p{\cal W}),
\end{array} 
\end{equation}
where  the $V_{AB}$ and ${\cal W}$ are scalars: $V_{AB}=v_{AB}$. Here, we used what the spin-isospin part of the wave function of the  fermion pair ${\Xi}{\Xi}$ is antisymmetric relatively to the permutation $P$ in Eq. (\ref{GrindEQ__1_}).

Let us assume that $V_{AA}=0$, then Eq. (\ref{eq:1av}) is reduced to a single equation:
\begin{equation} 
\label{eq:2av} 
\begin{array}{l} 
      (H_0+V_{AB}-E){\cal W}=-V_{AB} p{\cal W}.
\end{array} 
\end{equation}
This equation is similar to one obtained in Ref. \cite{GV16} for the $NN{\Xi}$, $(S,I)$=$(3/2,3/2)$ state within the integral Faddeev equations. However, the right hand side of Eq. (\ref{eq:2av}) has opposite sign. The restriction  $V_{AA}=0$ corresponds to the situation when  $NN$ potential can be neglected for the spin/isospin triplet $(S, I)=(1, 1)$  state.
The differential Eq. (\ref{eq:2av}) shows that the right hand side term is attractive (including  attractive ${\Xi}\alpha$ potential) and can give additional contribution into the binding energy  coming from the left hand side  term.
The corresponding term for the $NN{\Xi}$, $(S,I)$=$(3/2,3/2)$ state is repulsive due to symmetry  of $3/2$ spin/isospin basis functions relatively permutation of two identical particles  that holds the sign ''minus'' before the operator $P$ in Eq. (\ref{GrindEQ__1_}). 
The state $NN{\Xi}$, $(S,I)$=$(3/2,3/2)$ is unbound \cite{GV16}. 

\subsection{Interactions}
In this section we consider the two-body
interactions, which are the inputs to our present
study. To describe a nucleon-nucleon interaction, we use the semi-realistic Malfliet
and Tjon MT I-III \cite{MT} potential with the modification from Ref. \cite{MT1}.  
The MT I-III model has the Yukawa-type form:

$S=0$, $I=1$:  
$$
 V_{NN}(r)=(-513.968exp(-1.55r)+1438.72exp(-3.11r))/r  ,
$$

$S=1$, $I=0$:
$$
V_{NN}(r)=(-626.885exp(-1.55r)+1438.72exp(-3.11r))/r ,
$$
where the strength parameters are given in MeV and range parameters are given in fm$^{-1}$.
The parameters were chosen in Ref. \cite{MT} to reproduce the experimental data for $np$-scattering.
It has to be noted that we do  not use isospin formalism for the $nnp$ system. Thus, the protons and neutrons are not identical. The details of such treatment are presented in Ref. \cite{FSV16}.
To take into account that  the $nn$ interaction is not equivalent to $np$ interaction (that is known as the charge dependence of $NN$ interaction), we have made modification of the spin singlet $(S, I)=(0, 1)$  component of the MT I-III potential according Ref. \cite{FSV16} and have defined spin singlet $nn$ potential. 
The modification was performed by scaling strength parameter. The scaling parameter $\gamma$  is fixed as $\gamma$=0.975 to reproduce experimental $nn$ scattering length for which we used the value of -18.8~fm \cite{MO, chen}.
By this way,  we have obtained  three potentials $v^s_{nn}$, $v^s_{np}$ and $v^t_{np}$  needed  for Eq. (\ref{eq:1}).
Note that the MT I-III potential  is not defined for the  spin/isospin triplet $(S, I)=(1, 1)$ and singlet $(S, I)=(0, 0)$ states.  The corresponding potentials are taken to be equal zero. 

The $\Xi N$ and $\Xi \Xi$ potentials simulating the ESC08c Nijmegen model are written in the form \cite{GVV16}:

$S=0$, $I=1$:  
$$
 V_{\Xi N}(r)=(-290.0exp(-3.05r)+155.0exp(-1.6r))/r  ,
$$

$S=1$, $I=0$:
$$
V_{\Xi N}(r)=(-568.0exp(-4.56r)+425.0exp(-6.73r))/r ,
$$

$S=0$, $I=1$:  
$$
 V_{\Xi \Xi}(r)=(-155.0exp(-1.75r)+490.0exp(-5.6r))/r .
$$
The parameters of the potentials were fixed to reproduce the scattering lengths and effective radii given by the ESC08c Nijmegen model for the baryon-baryon interaction \cite{NRY15,RS16,RNY}.

For the $\Xi\alpha$ interaction we use the Isle-type potential \cite{Isle} which has the Gaussian form:
$$
V_{\Xi\alpha} (r) = 450.4 exp(-(r/1.269)^2)- 404.9 exp( -(r/1.41)^2).
$$
with parameters from Ref.  \cite{FSV08}. 

\section{Numerical results}
The ground state binding energies $B_3$ of the  $NNN$, $nnp$, $NN{\Xi}$,  $\Xi\Xi N$, $\Xi\Xi \alpha$   systems  were calculated using the  models suggested above. The numerical results are presented in Table \ref{t1}. 
For each system, we show the spin-isospin state $(S,I)$ and  two-body energies  $B_2(AA)$ and $B_2(AB)$ for AA and AB pairs.
 Additionally, we present the three-body binding energy calculated  under the condition $V_{AA}=0$.
Our results are compared with ones obtained  within the integral representation  of Refs. \cite{GVV16, G16}. 
One can see that results of  both approaches  are in the agreement with high accuracy.

\begin{table}[ht]
\caption{Binding energy $B_3$ (in MeV) calculated for various systems $AAB$ within the  differential (DFE) and integral Faddeev (IFE) equations. The  $B_3(V_{AA}=0)$ is
shown in  brackets.
Binding energy $B_2$ (in MeV)  for $AA$ and $AB$ pairs are also presented.
  Here $m_N$=938.91~MeV,  $m_\Xi$=1318.07~MeV, $m_\alpha$=3727.38~MeV.}
\centering
\label{t1}       
\begin{tabular}{lccccc}
\hline\noalign{\smallskip}
System & (Spin, Isospin) & $B_2(AA)$ & $B_2(AB)$ & $B_3$, DFE & $B_3$, IFE\cite{GVV16,G16} \\[3pt]
\hline
\noalign{\smallskip}
$NNN$&($1/2$, $1/2$) &2.23 &-- &8.58 \cite{FSV16} & -- \\  
$nnp$&($1/2$, --)&-- &2.23 &8.38\cite{FSV16} (3.40) &-- \\
\hline
$NN\Xi$ &($3/2$, $1/2$) & 2.23&1.67&17.205 (2.213)&  17.203 \\ 
$NN\Xi$ &($1/2$, $3/2$)& -- &  1.67&2.886 (1.785)& 2.8855 \\ \hline
$\Xi\Xi N$ &($1/2$, $3/2$) & -- & 1.67&4.512 (3.408) &4.5119 \\ 
$\Xi\Xi\alpha$ &($0$, $1$) & --& 2.09 &  7.635 (4.335)&  -- \\ 
\noalign{\smallskip}\hline
\end{tabular}
\end{table}

\begin{figure}[!ht]
\centering
\includegraphics[width=16pc]{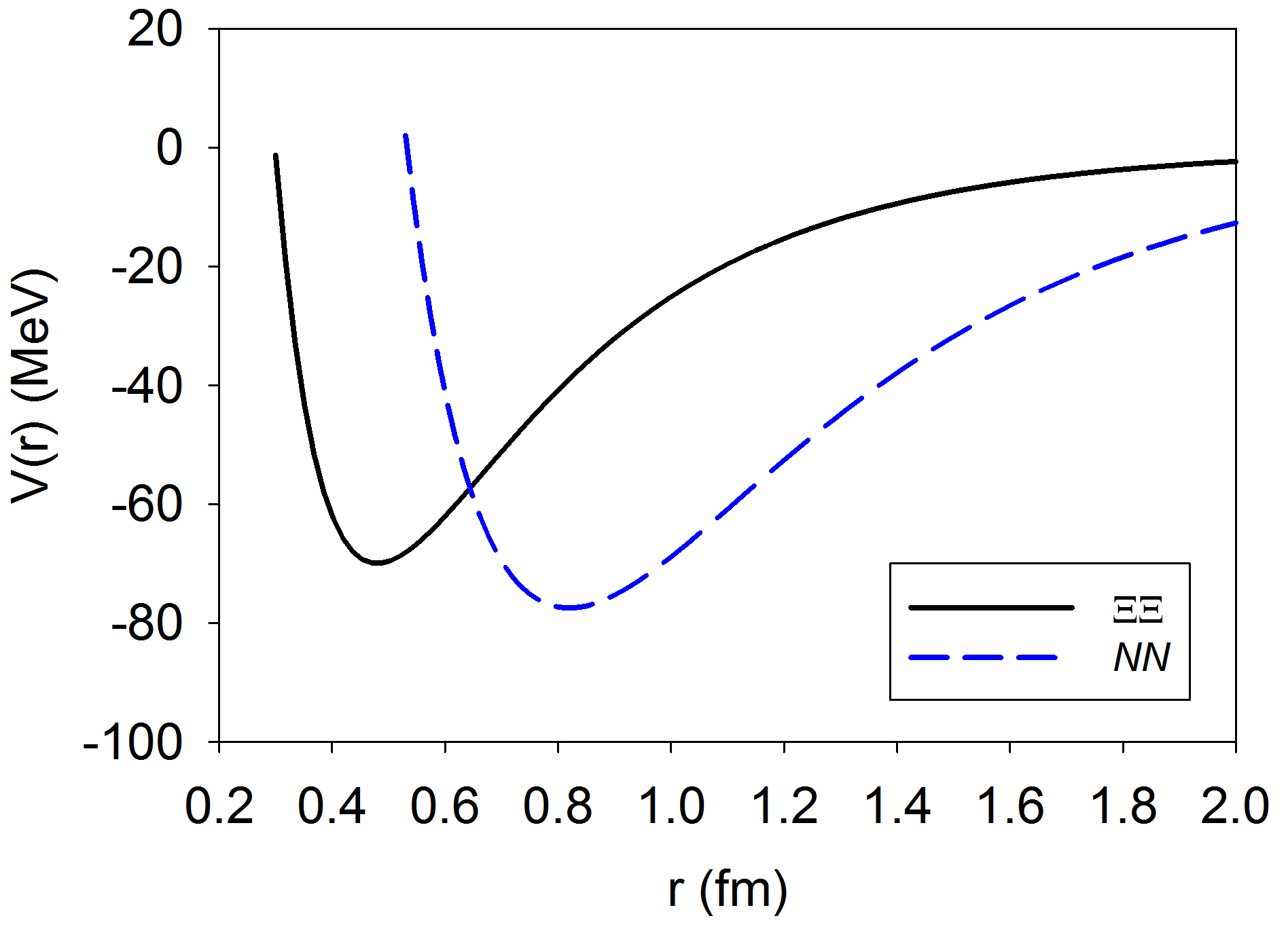}
\caption{\label{fig4} \small The $NN$  $(S,I)$=$(0,1)$ and  ${\Xi}{\Xi}$  $(S,I)$=$(0,1)$ potentials.
}
\end{figure}

\begin{figure}[!ht]
\centering
\includegraphics[width=29pc]{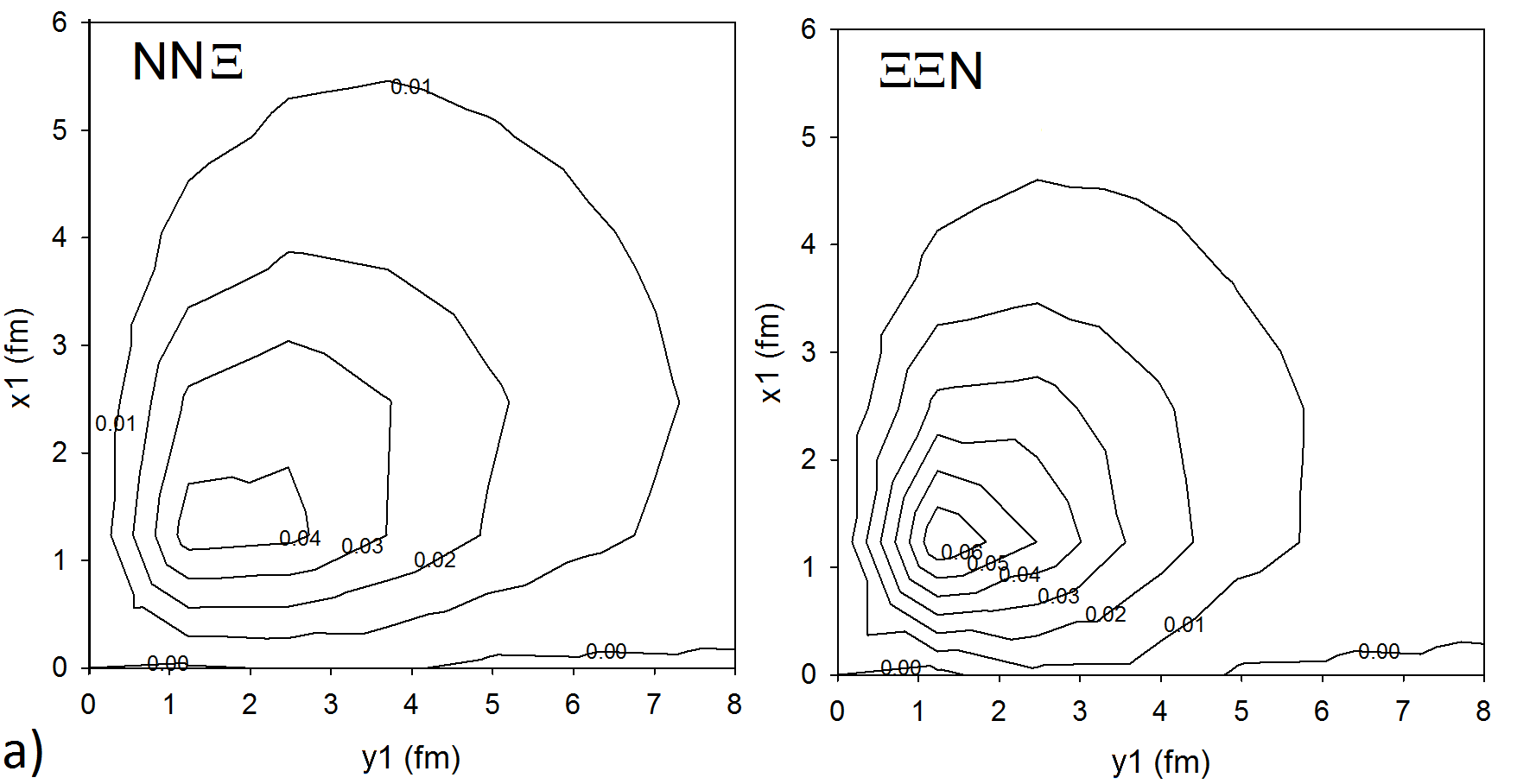}
\includegraphics[width=29pc]{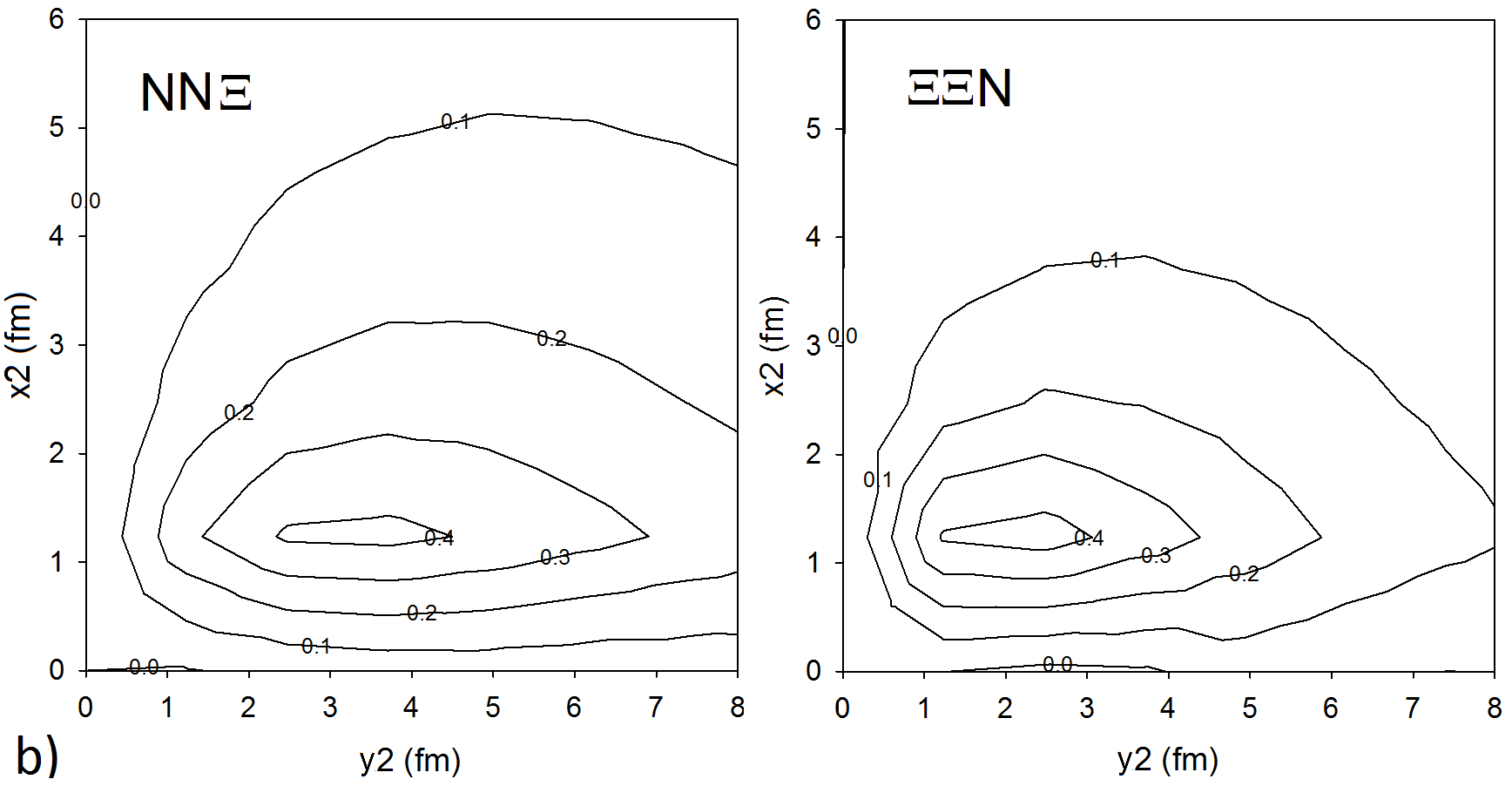}
\includegraphics[width=29pc]{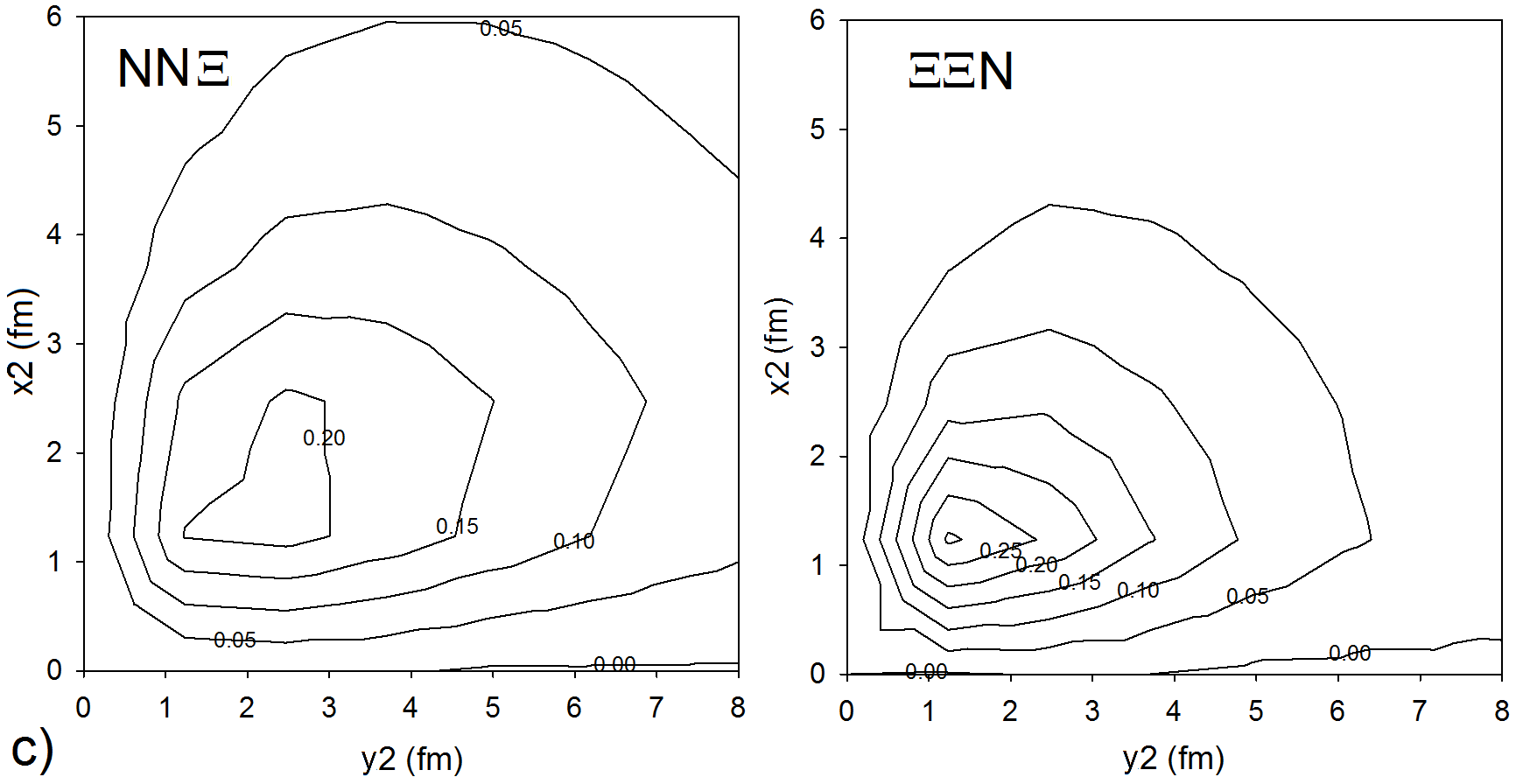}
\caption{\label{fig5} \small The contour plots of the  Faddeev components ${\cal U}$ a) and ${\cal W}$ b),c) for the $NN\Xi$ $(1/2,3/2)$ (Left)  and $\Xi\Xi N$ $(1/2,3/2)$ (Right) bound states. The Jacobi coordinates corresponding to the components ${\cal U}$ and ${\cal W}$ are presented as $x1,y1$ and $x2,y2$.
}
\end{figure}

The  "spin/isospin complication" \cite{FKSV17} of the Faddeev equations for the considered systems  is appeared by the matrix form of Eq. (\ref{w1}) and  leads to the following evaluation for the three-body binding energy of the $NN\Xi$ system
in the spin-isospin states $(S,I)$=$(3/2,1/2)$, $(1/2,3/2)$:
\begin{equation}\label{E3}
 B_2(AB)< B_3(V_{AA}=0)<2B_2(AB).
\end{equation}
The value of $B_3(V_{AA}=0)$ is restricted by 3.34~MeV. The similar result we have for the $nnp$ system. In this case, $B_3(V_{AA}=0)$ is restricted by 4.46~MeV.
In contrast,  the scalar form (\ref{eq:1av})  of Eq.  (\ref{eq:1})  for the case $\Xi\Xi\alpha$ $(S,I)$=$(0,1)$ leads to the relation:
\begin{equation}\label{E4}
B_3(V_{AA}=0)>2B_2(AB).
\end{equation}
This relation is known as the mass polarization effect which takes place when $m_B/m_A>1$ \cite{FG2002,H2002}. 
For the spin-isospin state $(S,I)$=$(0,1)$ of $\Xi\Xi\alpha$ system, the  mass polarization energy can be evaluated \cite{FKSV17}.
The contribution of this energy  $(B_3(V_{AA}=0)-2B_2(AB))/B_3(V_{AA}=0)$ in the three-body bound energy is equal 3.6\% that is  
compatible  with the values of 2\%-4\% \cite{H2002, FKSV17} for the similar nuclear system  $\Lambda\Lambda\alpha$.
The similarity takes place due to approximate equality of the masses of non-identical particles:  $m_B/m_A\sim 3$ for the $\Xi\Xi\alpha$ and  $m_B/m_A\sim 3$ for  $\Lambda\Lambda\alpha$. In the limit $m_B/m_A>>1$ the mass polarization effect can be neglected and $ B_3(V_{AA}=0)=2B_2(AB)$.
The case
$m_B/m_A<1$ is realized for the system $\Xi\Xi N$ $(1/2,3/2)$. 
There is no a simple relation between $B_3(V_{AA}=0)$ and $2B_2(AB)$ for the case.
One can define the incremental binding energy $\Delta B_{\Xi\Xi}$ for the system $\Xi\Xi N$ $(1/2,3/2)$  as $\Delta B_{\Xi\Xi}=B_3-2B_2(\Xi N)$ according 
 the analogue with the  $^6_{\Lambda\Lambda}$He hypernucleus. For  $^6_{\Lambda\Lambda}$He, the incremental binding energy is defined as $ \Delta B_{\Lambda\Lambda} = B_{\Lambda\Lambda}(^A_{\Lambda\Lambda}Z)-2B_\Lambda(^{(A-1)}_{\Lambda}Z)$ \cite{FG2002,H2002}. Calculating the incremental  energy, one can evaluate the strength  of the ${\Lambda\Lambda}$ interaction. For the system $\Xi\Xi N$ $(1/2,3/2)$, the energy includes significant 
 contribution of the mass polarization energy because $m_N/m_\Xi \sim 1$. Regardless that the  relation (\ref{E3}) is not satisfied, the more appropriate value for an evaluation of the strength  of the ${\Xi\Xi}$ interaction in $\Xi\Xi N$ $(1/2,3/2)$ is the value of $B_3-B_3(V_{\Xi\Xi}=0)$. The corresponding evaluation can be obtained from Table \ref{t1}. The ${\Xi\Xi}$ interaction is attractive in $\Xi\Xi N$ $(1/2,3/2)$. Analogically, the spin singlet $NN$ interaction is 
 also attractive in the $NN\Xi$ $(1/2,3/2)$ system. These attractive forces add about 1~Mev to the binding energies of the  mirror systems. Thus, the matrix elements $<\Psi|V_{AA}|\Psi>$ have the close values for the systems. It is possible, because 
 the $\Xi\Xi N$ system is more compact (larger $B_3$ value) and the $\Xi\Xi$ potential has a minimum closer to the origin as is shown in Fig. \ref{fig4}.  The Faddeev components $U$, ${\cal W}$ for the $NN\Xi$ $(1/2,3/2)$ and $\Xi\Xi N$ $(1/2,3/2)$ systems are presented in Fig. \ref{fig5}. From the figure, one can see  that the system  $\Xi\Xi N$ $(1/2,3/2)$ is  more compact than the $NN\Xi$ $(1/2,3/2)$ system. For  both systems, the
rearrangement channel   $A+(AB)$  dominates due to existence of the isospin singlet   $\Xi N$ bound state.

\begin{figure}[!t]
\centering
\includegraphics[width=33pc]{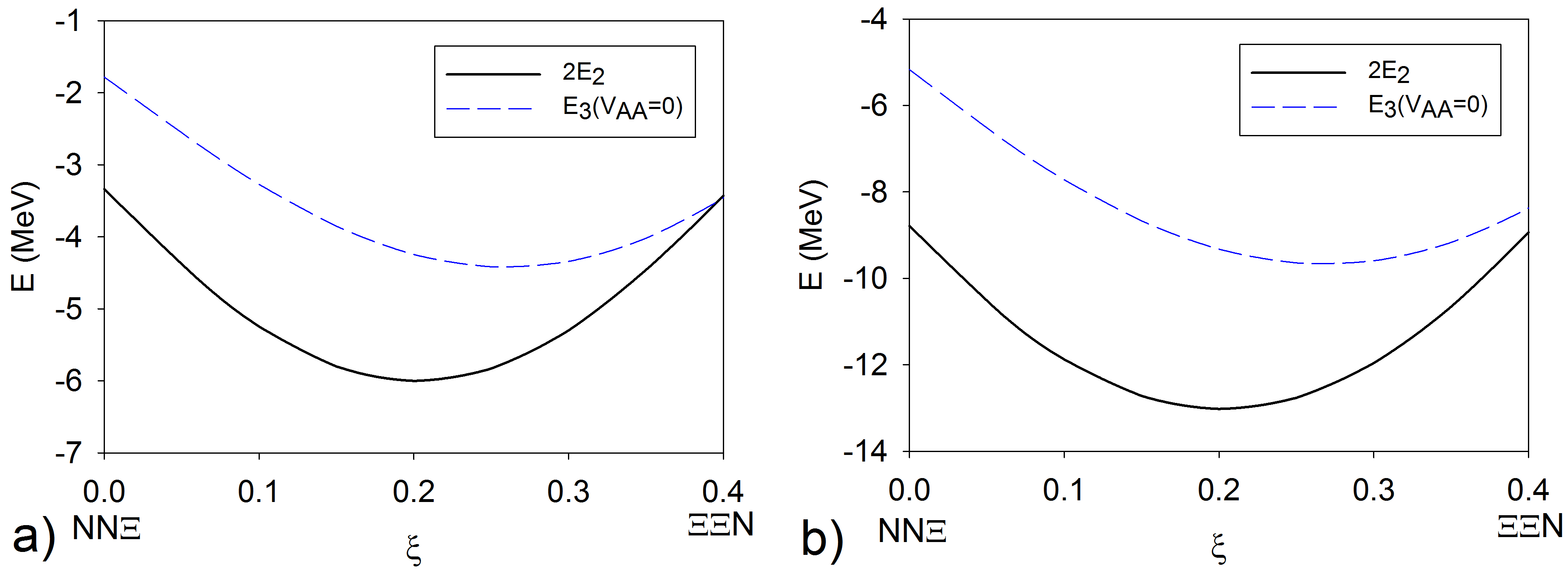}
\caption{\label{fig6} \small The transformation $NN\Xi$ $(1/2,3/2)$ to $\Xi\Xi N$ $(1/2,3/2)$ when $V_{AA}=0$. The   $2E_2$ (solid line) and $E_3(V_{AA}=0)$ (dashed line) as a function of $\xi$ are shown. The parameter $\xi$  is  related to the $NN\Xi$ $(1/2,3/2)$ system, when $\xi$=0, and - to the 
$\Xi\Xi N$ $(1/2,3/2)$ system, when $\xi$=0.4. a) The original spin triplet  $N\Xi$ potential is used. b) The spin triplet  $N\Xi$ potential is scaled by 1.05.
}
\end{figure}

The mirror $NN\Xi$ $(1/2,3/2)$ and $\Xi\Xi N$ $(1/2,3/2)$ systems under the condition  $V_{AA}=0$ can be  transformed  ''one into another'' by  changing the particle masses. The parameter $\xi \geq 0$  sets this transformation by the formula: $m^\xi_A= (1+\xi)m_A$,  $  m^\xi_B= (1-\xi m_A/m_B)m_B$.  The results of calculations for $2E_2$ and $E_3(V_{AA}=0)$ as a function of $\xi$ are shown in Fig. \ref{fig6}a). The transformation $NN\Xi$ $(1/2,3/2)$ to $\Xi\Xi N$ $(1/2,3/2)$ replaces the ratio $m_B/m_A>1$ to the ratio $m_B/m_A<1$. One can see that the relation (\ref{E3}) 
is well satisfied up to  $\xi$=0.2 when  $m^\xi_B/m^\xi_A \geq 1$. We  conclude that the relation (\ref{E3}) is not guaranteed when $m^\xi_B/m^\xi_A \ll 1$. The affect of  the $AB$ potential on the relation (\ref{E3})  is obvious. To show this we have repeated the calculations for more deep spin triplet $N\Xi$ potential. The potential has been scaled by the factor of 1.05. The result is shown in Fig. \ref{fig6}b). The relation (\ref{E3}) is satisfied for all possible values $\xi$ for this case.

It has to be noted that, as follows from Table \ref{t1}   for the  $NN{\Xi}$ $(3/2,1/2)$ state,   the three-body system having two bound subsystems has a deep bound state.
The value of this  $NN{\Xi}$ $(3/2,1/2)$ binding energy is related with two-body energies as  $B_3 >> 2B_2(AB)+B_2(AA)$. 
Obviously, the $V_{AA}$ potential plays a  key  role for formation of the bound state.
We  assume that it is a general property of such three-body systems. 


\section{Conclusions}
We studied the  hypernuclear system $NN\Xi $ (and $\Xi\Xi N$) based on the configuration-space Faddeev equations.
The baryon-baryon potential of ESC08c model, which generates  the $\Xi N$ $(S,I)=(1,1$) $s$-wave bound state, results in  the stable states for these three-body systems. The stability relatively $N\Xi \to \Lambda\Lambda$ conversion
is provided by fixing the states with maximal isospin.  
Our results and ones obtained  within the integral Faddeev equation formalism \cite{GVV16, G16} are in agreement with high accuracy. 
Additionally, we have calculated the binding energy of the $\Xi\Xi \alpha$  $(S,I)=(0,1)$ state.
The relations between $B_2$ and $B_3(V_{AA}=0)$ were proposed for the "spin/isospin complicated" and ''scalar"  states. The corresponding relations are significantly different.


\end{document}